\documentclass[usenatbib,useAMS, usegraphicx]{mn2e}

\newcommand{\hp}{H\ensuremath{_{3}^{+}}}
\newcommand{\ddp}{D\ensuremath{_{3}^{+}}}
\newcommand{\hdp}{H\ensuremath{_{2}}D\ensuremath{^{+}}}
\newcommand{\dhp}{D\ensuremath{_{2}}H\ensuremath{^{+}}}
\newcommand{\hh}{H\ensuremath{_2}}
\newcommand{\dd}{D\ensuremath{_2}}
\newcommand{\cm}{cm\ensuremath{^{-1}}}
\newcommand{\et}{et al.}

\begin{document}
\title{Deuterated hydrogen chemistry: Partition functions, Equilibrium constants and Transition intensities for the \hp\ system}

\author[Jayesh Ramanlal and Jonathan Tennyson]{Jayesh Ramanlal and Jonathan Tennyson \\
Department of Physics and Astronomy, University College London,
London, WC1E 6BT}
\maketitle

\begin{abstract}
\hp\ and the deuterated isotopomers are thought to play an important role in interstellar chemistry. The partition functions of \hp\, \dhp\ and \ddp\ are calculated to a temperature of 800 K by explicitly summing the {\it ab initio} determined rotation-vibration energy levels of the respective species. These partition functions are used to calculate the equilibrium constants for nine important reactions in the interstellar medium involving \hp\ and its deuterated isotopomers. These equilibrium constants are compared to previously determined experimental and theoretical values. The Einstein A coefficients for the strongest dipole transitions are also calculated. 
\end{abstract}


\section{Introduction}

Deuterium chemistry in the interstellar medium has had renewed
interest of late, this is due in part to recent observations of
multiply deuterated species in the interstellar medium
\citep{lcc01,c02,vpy04}. The cosmic abundance of deuterium with
respect to hydrogen at low temperature is approximately
1.4$\times10^{-5}$ in the solar neighbourhood, \citep{laj99}, but a
much higher ratio is observed between molecules and their deuterium
baring isotopomers in some environments. Molecules containing
deuterium can become highly fractionated in the gas phase at low
temperatures. This effect is thought to be primarily through reactions
with \hdp. \hdp was first detected by \cite{std99} in the interstellar
medium and more recently by \cite{ctc03}. However modelling of
interstellar deuterium chemistry by \cite{rhm03} suggest that that all
the deuterated \hp\ isotopomers have an effect on fractionation. That
is the inclusion of \ddp~and
\dhp~enhances fractionation significantly.

The other mechanism by which deuterium fractionation can occur is via active
grain chemistry. That is to say that molecules which are stuck onto
grains are deuterated by deuterium which accrete onto the grain
mantle. \cite{blc03} suggest that in prestellar
cores, deuteration increases with increasing CO depletion. This would
indicate that deuteration is primarily the product of gas phase
chemistry as CO depletion leads to an increase in [\hdp]/[\hp]
abundances \citep{dl84}.

Hydrogenic gas phase reactions involving \hp~and the deuterated
isotopomers that are regarded to be significant in gas phase
deuteration are tabulated in table~\ref{tab:enthal}. Within the
thermodynamic equilibrium regime, the equilibrium constants, K, of
these reactions can be calculated from the partition functions of the
reactant and product species. It is not possibly to separate the
forward and backward rates in these calculations thus comparisons can
only been made to the ratio of the forward k$_f$, and backward rates
k$_b$.
\begin{equation}
	K = \frac{k_f}{k_b}
\end{equation}

Experiments measuring both the forward and backward rates for the
reactions of interest have been conducted by \cite{as81}, \cite{gas92}
and most recently by \cite{ghr02}. Both Adams and Smith and Giles
\et~used a variable temperature selected ion flow arrangement
\citep{saa82}, while Gerlich \et~used a low temperature multipole ion trap.

Previously theoretical studies have concentrated on the
\hp~partition functions and associated equilibrium constants for
reactions of interest; most notably in \cite{jt113} and
\cite{jt169}. Sidhu \et\ calculated the partition function
of \hp~ and \hdp~ to a temperature of 2800 K. Sidhu \et\ also
calculated a number of equilibrium constants. Neale \et\ calculated
the high temperature \hp~ partition function which extended to
10000K. We recalculate the \hp partition function in order to compare
with these previous calculations and verify our
methodology. \cite{h82} calculated partition functions using what was
then rather limited experimental and calculated molecular data,
from these equilibrium constants were determined in a similar manner
to this work.

Many attempts have been made to observe the deuterated species of
\hp. To this end the Einstein A coefficients of the dipole transitions
have been calculated. The strongest transitions have been selected and
are tabulated here.

\section{Method}  

\subsection{Partition function}
The partition function of \hdp\ was taken from Sidhu \et\ The
internal partition functions, z$_{int}$ for \hp, \ddp\ and \dhp\ were
computed by explicitly summing the series:
\begin{equation}
\label{eq:part_sum}
z_{int}=\sum_{i} (2J+1)g_i \exp\left(-\frac{c_{2}E_{i}}{T}\right)
\end{equation}
where J is the rotational quantum number, $g_i$ is the nuclear spin
degeneracy factor for state {\it i}, c$_2$ is the second radiation
constant and E$_i$ is the associated energy level relative to the J=0
ground state in \cm. No distinction was made between
rotational and vibrational energy levels. \hp\ has only one bound
electronic bound state, thus there was no electronic contribution to
the partition functions. All the \hp\ energy levels were taken relative
to the J=0 vibrational ground state.

The full partition function, z$_{tot}$ can be written as
\begin{equation}
z_{tot}=z^{trans}z_{int}
\end{equation}
where z$_{int}$ is the internal partition function. z$^{trans}$ is the
translational contribution to the partition function which can be
estimated using the perfect gas model. As all the reactions considered
in this work conserve the number of particles in the system. the ratio
of their translational partition functions is given by a simple mass
factor \citep{h67}
\begin{equation}
\frac{z_{C}^{trans}z_{D}^{trans}}{z_{A}^{trans}z_{B}^{trans}}
=\left( \frac{m_{C}m_{D}}{m_{A}m_{B}} \right)^{3/2}
\end{equation}
where m$_X$ is the mass of species X. 

\subsection{Equilibrium constants}
For reaction
\begin{equation}
	A+B \rightleftharpoons C+D
\end{equation} 
the temperature dependent equilibrium constant, K(T), was calculate
using the following,
\begin{equation}
	K=\frac{z^{C}_{tot}z^{D}_{tot}}{z^{A}_{tot}z^{B}_{tot}} 
	\exp\left(-\frac{U}{kT}\right)
\end{equation}
where z$_{tot}$ is the partition function incorporating translational motion
and U is the enthalpy of the reaction. The enthalpy of the reaction
was calculated using
\begin{equation}
U=E_{0}^{C}+E_{0}^{D}-E_{0}^{A}-E_{0}^{B}
\end{equation}
where E$_{0}^{X}$ is the zero point energy of species X as measured on
an absolute energy scale. Thus the enthalpies for reactions (b) to (i)
were calculated in this way. The diatomic zero point energies were
calculated using the constants of \cite{hh79} (table~\ref{tab:consts})
and equation~\ref{eq:vib_terms}. The zero point energies of \hp~ and
isotopomers were taken from \cite{jt318}. The ground state for \hp~is
forbidden by the Pauli principle; The lowest state, J=1, K=1, lies
some 64.123 \cm\ above this ground state \citep{jt318}. Thus for \hp~the
so called ``rotational zero point energy'' was used, which is
4425.823 \cm. For reaction (a) the difference in ionisation energy
between H and D was taken to be 46.4 K \citep{k87}.

\begin{table}
\caption{Calculated enthalpies for the reactions of interest
\label{tab:enthal}
}
\begin{tabular}{lll}
\hline \hline
 &  Reaction & Enthalpy / K \\
\hline
(a) & \hp + D $\rightarrow$ \hdp + H & $-$597.8  \\

(b) & \hp + HD $\rightarrow$ \hdp + \hh &  $-$231.8  \\

(c) & \hdp + HD $\rightarrow$ \dhp + \hh & $-$187.2  \\

(d) & \dhp + HD $\rightarrow$ \ddp + \hh & $-$233.8  \\

(e) & \hp + \dd $\rightarrow$ \hdp + HD & $-$153.0  \\

(f) & \hp + \dd $\rightarrow$ \dhp + \hh & $-$340.2  \\

(g) & \hdp + \dd $\rightarrow$ \dhp + HD & $-$108.4  \\

(h) & \hdp + \dd $\rightarrow$ \ddp + \hh & $-$342.2  \\

(i) &\dhp  + \dd $\rightarrow$ \ddp + HD & $-$155.0  \\
\hline
\end{tabular}

\end{table}

The diatomic partition functions needed for the equilibrium constants
were calculated using the formulae given below.
\begin{equation}
z=\sum_{\nu,J}(2J+1)g_{J}\exp\left(-\frac{F_{\nu}+G_{\nu}-G_{0}}{kT}\right)
\end{equation}
where
\begin{equation}
	F_{\nu}=B_{\nu}J(J+1)-D_{e}J^2(J+1)^2,
\end{equation}
\begin{equation}
	B_{v}=B_{e}=\alpha_{e}(\nu+{\textstyle\frac{1}{2}}),
\end{equation}
\begin{equation}
\label{eq:vib_terms}
	G_{\nu}=\omega_{e}(\nu+{\textstyle\frac{1}{2}})-
	\omega_{e}x_{e}(\nu+{\textstyle\frac{1}{2}} )^2 
\end{equation}
The constants used were taken from \cite{hh79} and
are tabulated in table~\ref{tab:consts}.

\begin{table}
\caption{Constants in \cm\  used to calculate diatomic partition functions, taken from ~\citet{hh79}}
\label{tab:consts}
\begin{tabular}{llll}
\hline \hline
   & H$_2$ & D$_2$ & HD \\
\hline
B$_e$ 		& 60.853 & 30.443 & 45.655 \\
$\alpha_e$ 	& 3.06 	& 1.0786 & 1.986 \\
D$_e$ 		& 0.0471 & 0.01141 & 0.02605 \\
$\omega_e$ 	& 4401.21 & 3115.50 & 3813.15 \\
$\omega_{e}x_e$ & 121.33 & 61.82 & 91.65 \\
g$_e$ 		& 1 & 6 & 6 \\
g$_o$ 		& 3 & 3 & 6 \\ 
D$_o^o$		& 36118  & 36406 & 36749 \\
$\nu_{max}$	& 14	 & 20    & 16 \\
 
\hline
\end{tabular}
\end{table}

\subsection{Energy levels}
\label{sec:energy_levels}
Energy levels for \hp, \ddp\ and \dhp\ were determined using the DVR3D
program suite of \cite{jt334} and the ultra-high accuracy {\it ab
initio} procedure of \cite{jt236} which was based on the electronic
structure calculations of \cite{crj98}. The parameters used for the
DVR3D program suite are the same as those outlined in \cite{jt236}.A
total of 40$\times$(J+1) levels were computed for each J, up to
J=14. This procedure gave at least 19119 rotation vibration energy
levels for each molecule and ensured that all energy levels up to
10000~\cm~were included.

\dhp\ has C$_{2\nu}$ symmetry, this symmetry is fully represent in the DVR3D 
program by the parity of the basis, even and odd; which means that
energies with even (g$_{e}=$3) and odd (g$_{o}=$6) parity are easily
easily identified. \hp~ and
\ddp~ have D$_{3h}$ symmetry, this symmetry is not fully represented
by the DVR3D program. Thus energies with E, A$_1$ and A$_2$ cannot be
so easily identified. The A$_1$ and A$_2$ states are represented by
DVR3D and have even and odd basis parity respectively. The E symmetry
energies are determined by the fact that they are degenerate across
even and odd basis by parities. Thus the E symmetry states can be
identified by examining by hand the complete list of energy levels
for both even and odd basis parities. For \hp\ the nuclear
degeneracy factors are 2, 0 and 4 for E, A$_1$ and A$_2$
respectively. For \ddp~ the the nuclear degeneracy factors are 8, 10
and 1 for E, A$_1$ and A$_2$ respectively. 

There has been some confusion in the conventions used for nuclear spin
degeneracy factor, g. In general astronomers use the nuclear spin
degeneracy divided by its value for the dissociated atoms, while
physicists use the nuclear spin degeneracy itself. Thus for example
astronomers would take g for the E states of \ddp\ to be 8/27 while we
have used 8. Both conventions are equally valid; however it is
important to be consistent when calculating equilibrium constants that
the same convention has been used in calculating the partition
functions for all the species in the reaction. Throughout this work
the ``physicists'' convention is used.

\section{Results and Discussion}

\subsection{Partition function}
Table~\ref{tab:part_sum} present the values obtained by the explicit
summation of equation~\ref{eq:part_sum}. It was found that at a
temperature of 800 K the inclusion of the J=14 energy levels only
contributed only 0.02\%, 0.77\% and 0.35\% to the internal partition
functions for \hp\ and \ddp~ and \dhp\ respectively. This the partition
functions are valid up to a temperature of 800 K.
\begin{table}
\caption{Calculated internal partition functions as a function of a temperature. Powers of ten given in parenthesis
\label{tab:part_sum}}
\begin{tabular}{rrrr}
\hline
\hline
      T(K) & H$_{3}^{+}$ & D$_{3}^{+}$ & D$_{2}$H$^{+}$ \\
\hline
         5 & 5.8325($-$08) &  10.0022   &     6.0008 \\
			      	 
        10 & 6.3491($-$04) &  10.2351   &     6.1281 \\
			      	 
        20 &        0.0826 &  12.5729   &     7.8853 \\
			      	 
        30 &        0.4654 &  16.3763   &    11.1196 \\
	         	      	 
        40 &        1.1481 &  21.1870   &    15.0916 \\
	         	      	 
        50 &        2.0197 &  26.9604   &    19.5932 \\
	         	      	 
        60 &        2.9948 &  33.6025   &    24.5777 \\
	         	      	 
        70 &        4.0243 &  40.9934   &    30.0255 \\
	         	      	 
        80 &        5.0826 &  49.0277   &    35.9154 \\
	         	      	 
        90 &        6.1579 &  57.6247   &    42.2237 \\
	         	      	 
       100 &        7.2457 &  66.7263   &    48.9256 \\
	         	      	 
       150 &       12.9039 &  118.6694  &    87.6098 \\
	         	      	 
       200 &       19.0975 &  179.7917  &  133.6128 \\
	         	      	 
       300 &       33.3860 &  325.2799  &  243.7791 \\
	         	      	 
       400 &       50.0599 &  498.7918  &  375.2491 \\
	         	      	 
       500 &       68.9696 &  701.1151  &  527.2953 \\
	         	      	 
       600 &       90.1816 &  936.5854  &  701.5534 \\
	         	      	 
       700 &      113.9581 &  1211.2871 &  901.0749 \\
	         	      	 
       800 &      140.6995 &  1531.8784 &  1129.5545 \\
\hline
\hline
\end{tabular}  
\end{table}

\begin{figure*}
        \centerline{\includegraphics[scale=0.65,angle=-90]{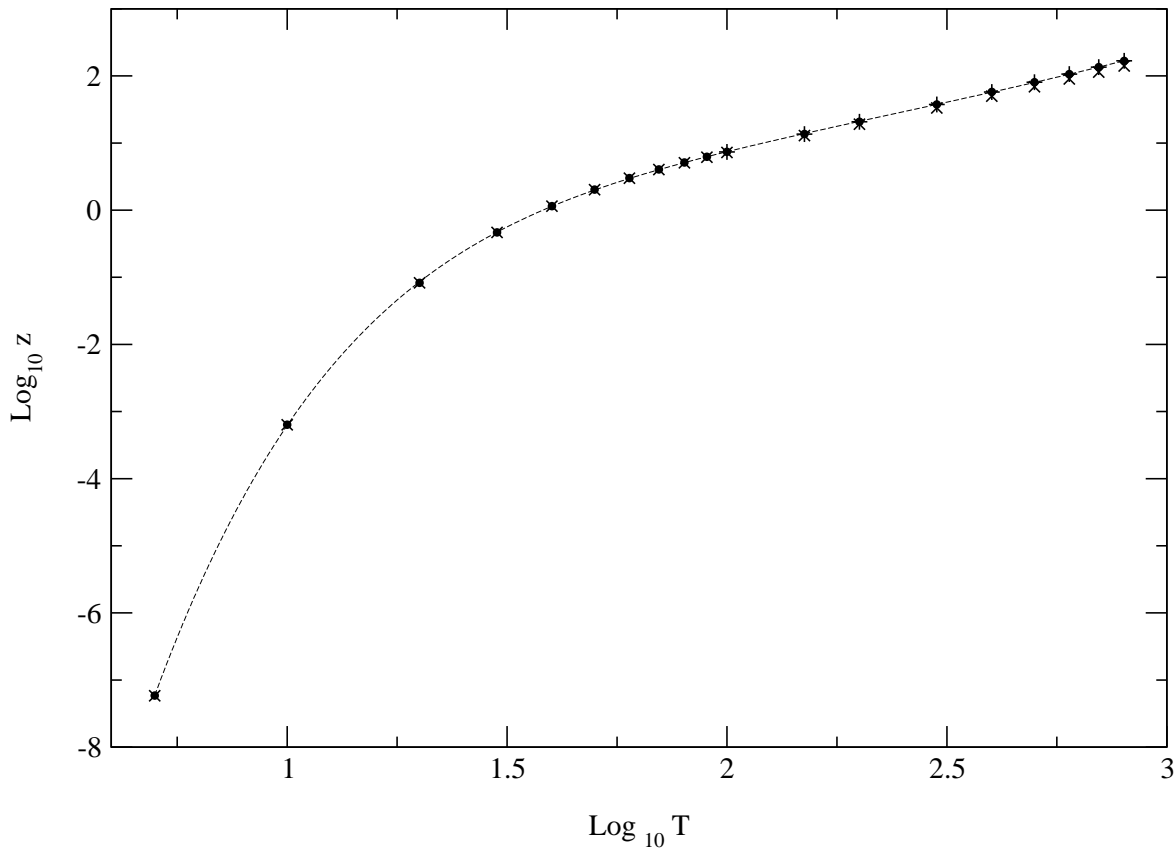}}
        \caption{\hp~partition function, {\it z} as a function of
        temperature, T. {\it Crosses}, this calculation; {\it dashed
        curve}, fit of equation \ref{eq:fit} to our calculated data;
        {\it Circles}, calculation of \citet{jt113}; {\it Pluses},
        calculation of \citet{jt169}.}
\label{fig:part_com}
\end{figure*}

\begin{figure*}
        \centerline{\includegraphics[scale=0.65,angle=-90]{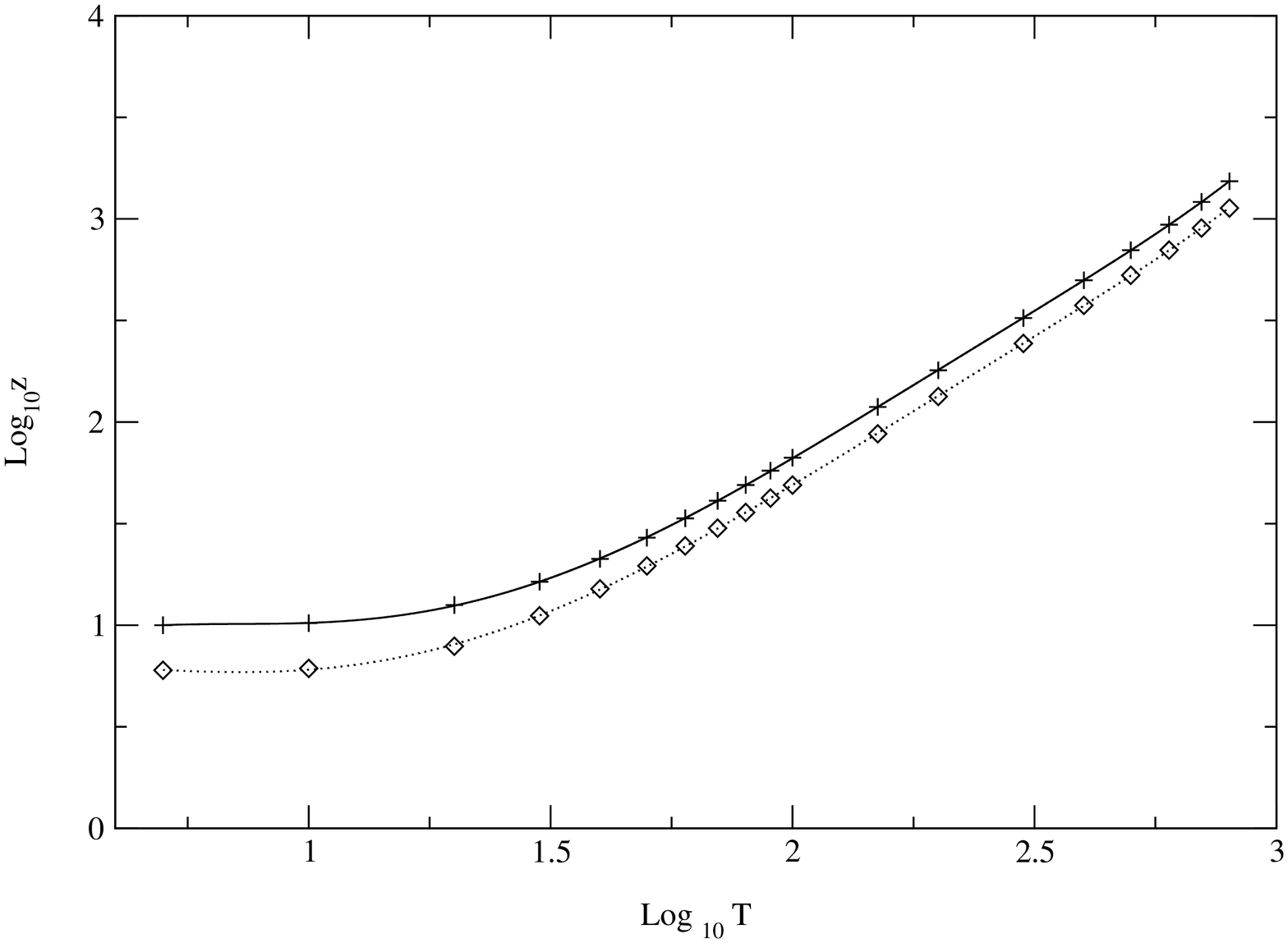}}
        \caption{\dhp~and \ddp~partition functions, {\it z}, as a function of temperature, T. {\it Diamonds}, \dhp~calculation; {\it dashed curve}, fit of equation \ref{eq:fit} to \dhp~data; {\it Crosses}, \ddp~calculation; {\it solid curve}, fit of equation \ref{eq:fit} to \ddp~data.}
\label{fig:part}
\end{figure*}

Comparing the \hp\ partition functions of this work and that of Sidhu
\et (figure~\ref{fig:part_com}) we see that the two works are
in good agreement. There is some minor disagreement at higher
temperatures where in any case the much more comprehensive partition
function of \cite{jt169} should be used.

The partition functions have been fitted to the standard formula, see
\cite{st84}, in the temperature range 5 K to 800 K using the data in
table~\ref{tab:part_sum}. The coefficients a$_n$ for
\hp~, \ddp~ and \dhp~ are tabulated in table~\ref{tab:part_fit}.

\begin{equation}
\label{eq:fit}
\log_{10}(z)=\sum_{n=0}^{6} a_n (\log_{10} T)^n
\end{equation}

Our fit is never more than 1.35\%, 0.78\%, and 1.22\% from the calculated
values of {\it z} for \hp, \ddp and \dhp respectively.
\begin{table}
\begin{tabular}{lrrr}
\hline \hline
	& \hp	& \ddp	& \dhp	\\
\hline
a$_0$ &  $-$35.10102 &  $-$0.388363   &  	$-$0.975341 \\
a$_1$ &  72.2463   &  5.65495     &  	7.92203   \\
a$_2$ &  $-$66.3543  &  $-$8.53925    &	$-$13.5834  \\
a$_3$ &  35.3938   &  5.83071     &	11.0576   \\
a$_4$ &  $-$11.3756  &  $-$1.74965    &	$-$4.45541  \\
a$_5$ &  2.06118   &  0.205985    &	0.890251  \\
a$_6$ &  $-$0.160957 &  $-$0.00289176 &	$-$0.0704028\\
\hline
\end{tabular}
\caption{Fitting coefficients for the polynomial fit (equation~\ref{eq:fit}) of the partition functions in the temperature range 5 K to 800 K.
\label{tab:part_fit}}
\end{table}

\subsection{Equilibrium Constants}
Table~\ref{tab:equlib_cons} gives the logarithm equilibrium constants,
K, for the reactions tabulated in table~\ref{tab:enthal} as a function
of temperature. The equilibrium constants were calculated using the
partition functions previously discussed. The \hdp partition functions
were taken from \cite{jt113}.

\begin{table*}
{\begin{minipage}{210mm}
\caption{Equilibrium constants, K, for the reactions given in table~\ref{tab:enthal}, as a function of temperature. The first line corresponds to reactants while the second line corresponds to products. 
\label{tab:equlib_cons}}
\begin{tabular}{llllllllll}
\hline \hline
     T (K)&  \hp+D & \hp+HD & \hdp+HD & \dhp+HD & \hp+\dd & \hp+\dd & \hdp+\dd & \hdp+\dd & \dhp+\dd \\
       &\hdp+H    & \hdp+\hh & \dhp+\hh & \ddp+\hh & \hdp+HD & \dhp+\hh & \dhp+HD & \ddp+\hh & \ddp+HD \\ \hline
         5 &   59.1951 &    26.5161 &    16.1410 &  19.6059 &    20.5256 &    36.6666 &    10.1505 & 29.7564   & 13.6154  \\
        10 &   29.2007 &    12.4148 &     8.0189 &  9.4544  &     9.8455 &    17.8644 &     5.4496 & 14.9040   &  6.8851   \\
        20 &   14.1069 &     5.2661 &     4.0622 &  4.3690  &     4.4023 &     8.4645 &     3.1984 &  7.5674   &  3.5052   \\
        30 &    9.3591 &     3.1640 &     2.5237 &  2.6552  &     2.8620 &     5.3857 &     2.2218 &  4.8770   &  2.3533   \\
        40 &    7.0102 &     2.1462 &     1.7791 &  1.8049  &     2.1178 &     3.8969 &     1.7507 &  3.5556   &  1.7765   \\
        50 &    5.6306 &     1.5816 &     1.3426 &  1.3131  &     1.7047 &     3.0473 &     1.4657 &  2.7787   &  1.4361   \\
        60 &    4.7257 &     1.2337 &     1.0659 &  0.9998  &     1.4456 &     2.5115 &     1.2777 &  2.2776   &  1.2117   \\
        70 &    4.0858 &     0.9997 &     0.8802 &  0.7841  &     1.2668 &     2.1471 &     1.1473 &  1.9314   &  1.0512   \\
        80 &    3.6116 &     0.8337 &     0.7465 &  0.6256  &     1.1377 &     1.8842 &     1.0505 &  1.6760   &  0.9296   \\
        90 &    3.2462 &     0.7093 &     0.6456 &  0.5030  &     1.0398 &     1.6854 &     0.9761 &  1.4792   &  0.8335   \\
       100 &    2.9563 &     0.6122 &     0.5664 &  0.4045  &     0.9632 &     1.5296 &     0.9174 &  1.3219   &  0.7555   \\
       150 &    2.1009 &     0.3270 &     0.3283 &  0.0973  &     0.7441 &     1.0724 &     0.7454 &  0.8426   &  0.5143   \\
       200 &    1.6813 &     0.1816 &     0.2021 & -0.0694  &     0.6415 &     0.8436 &     0.6620 &  0.5925   &  0.3904   \\
       300 &    1.2660 &     0.0311 &     0.0676 & -0.2472  &     0.5457 &     0.6133 &     0.5822 &  0.3350   &  0.2674   \\
       400 &    1.0602 &    -0.0445 &    -0.0014 & -0.3386  &     0.5062 &     0.5048 &     0.5493 &  0.2108   &  0.2122   \\
       500 &    0.9374 &    -0.0893 &    -0.0421 & -0.3927  &     0.4914 &     0.4493 &     0.5387 &  0.1460   &  0.1880   \\
       600 &    0.8569 &    -0.1179 &    -0.0687 & -0.4274  &     0.4904 &     0.4217 &     0.5397 &  0.1123   &  0.1810   \\
       700 &    0.7999 &    -0.1380 &    -0.0866 & -0.4507  &     0.4967 &     0.4101 &     0.5481 &  0.0974   &  0.1840   \\
       800 &    0.7579 &    -0.1523 &    -0.0997 & -0.4668  &     0.5076 &     0.4079 &     0.5602 &  0.0934   &  0.1931   \\
\hline
\end{tabular}
\end{minipage}}
\end{table*}

There have been a few experiments where both the forward and backward
reaction rates of interest have been measured; so that the equilibrium
constant may be deduced for comparison. A comparison of the
experimental data available to date is given in tables
\ref{tab:gas92_com}, \ref{tab:as81_com} and
\ref{tab:ghr02_com}.  Our calculations generally show approximate agreement
with the experimental data. The notable exception is the recent
experiment of \cite{ghr02} for the
\hp + HD$\rightarrow$\hdp +\hh, which disagrees with our calculations
by 12 orders of magnitude. Gerlich \et~ measured the forward and
backward rates at a low temperature, 10 K, using a low temperature
multipole ion trap. There have been no other experiments carried out
at this low temperature, thus no direct experimental comparison can be
made. Smith and Adams using standard extrapolation give an equilibrium
constant of 1.7$\times$10$^{+9}$ at 10 K, which is in better agreement
to our own result. Most models use the the equilibrium
constants of \cite{as81}. The previously calculated value of Sidhu
\et~ used an enthalpy of 139.5 K, this enthalpy does
not take into account the previously discussed rotational zero point
energy. Thus if an enthalpy of 231.8 K is used then an equilibrium
constant of 7.1$\times$10$^{+12}$, which is more consistent to our own
is obtained.

A comparison of our equilibrium constants with those of \cite{gas92}
for the reactions of interest are shown in table
\ref{tab:gas92_com}. Giles \et\ give relative errors of $\pm$ 15\% for the
equilibrium constants. This relates to a much lower absolute value on
the measured rate; Giles \et\ state that certain systematic errors
will cancel when taking the ratio of rates, thus producing lower error
bounds. This may be rather optimistic. In general there is better
agreement between our calculation and the experiment of Giles \et\ at
the higher temperature of 300 K. This is most likely due to the less
demanding nature of measuring the reaction rates as higher
temperature. Giles \et\ also calculate the equilibrium constants by
calculating the partition functions of the reactant and product
species, and also the enthalpy change for the reaction. These
partition functions were obtained by explicitly summing the energy
levels (as equation \ref{eq:part_sum}) as given by using the rigid
rotor approximation and the relevant experimentally determined
rotational constants. The rigid rotor approximation is problematic for
the \hp\ system and definitely inferior to our own {\it ab initio}
energy level calculations. However we are generally in better
agreement with Giles
\et\ theoretical equilibrium constants than their experimentally
derived equilibrium constants.

A comparison of equilibrium constants for the reaction \hp + HD
$\rightarrow$ \hdp + \hh\ for a number of temperatures to those of
\cite{as81} and \cite{h82} are given in table \ref{tab:as81_com}. Adams
and Smith estimate their errors on the reaction rates to be $\pm$20\%,
this gives an error on the equilibrium constants of $\pm$30\%. We are
generally in good agreement with Adams and Smith. \cite{h82}
calculated the reaction constants using his calculated partition
functions and reaction enthalpy. The partition functions for \hp\ and
\hdp are determined from explicitly doing the sum as in equation
\ref{eq:part_sum}. The energy levels are found by using the
spectroscopic constants of \cite{o80,o81} and \cite{car80}
respectively. The constants of \cite{hh79} were used to calculate the
partition functions of HD and \hh. Our calculations are in good
agreement with those of \cite{h82}.

\begin{table*}
{\begin{minipage}{420mm}
\caption{A comparison of Equilibrium constants with \citet{gas92}.}
\label{tab:gas92_com}
\begin{tabular}{llllllll}
\hline
\hline
       & {\bf 80K} & & & & {\bf 300K} & & \\
\cline{2-4} \cline{6-8}
       & Giles \et\ & Giles \et\ & This work & & Giles \et\ & Giles \et\ & This work \\
       & (Expt)        & (Theory)      &           & & (Expt)        &  (Theory)    &            \\
\hline
\hp + HD $\rightarrow$ \hdp + \hh   &   3.8 ($\pm$0.6) & 6.6 &  6.82 & &   1.80 ($\pm$0.3) & 2.0 &  1.07 \\
						      		    
\hdp + HD $\rightarrow$ \dhp + \hh  &   1.7 ($\pm$0.3) & 3.6 &  5.58 & &  0.80 ($\pm$0.1) & 0.8 &  1.17 \\
						      		    
\dhp + HD $\rightarrow$ \ddp + \hh  &   1.8 ($\pm$0.3) & 2.0 &  4.22 & &  0.40 ($\pm$0.1) & 0.3 &  1.00 \\
						      		    
\hp + \dd $\rightarrow$ \hdp + HD   &   8.2 ($\pm$1.2) & 13.2 & 13.73 & &  5.20 ($\pm$0.8) & 6.7  &  3.51 \\
						      		    
\hp + \dd $\rightarrow$ \dhp + \hh  &   9.2 ($\pm$1.4) & 48.3 & 76.59 & &  5.30 ($\pm$0.8) & 5.1 &  4.10 \\
						      		    
\hdp + \dd $\rightarrow$ \dhp + HD  &   4.4 ($\pm$0.7) & 7.2 & 11.23 & &  1.90 ($\pm$0.3) & 2.5 &  3.82 \\
						      		    
\hdp + \dd $\rightarrow$ \ddp + \hh &   4.4 ($\pm$0.7) & 14.5 & 47.43 & &  0.70 ($\pm$0.1) & 0.8 &  2.16 \\
						      		    
\dhp  + \dd $\rightarrow$ \ddp + HD &   1.7 ($\pm$0.3) & 4.0 &  8.50 & &  0.80 ($\pm$0.1) & 1.0 &  1.85 \\
\hline
\end{tabular}  
\end{minipage}}
\end{table*}

\begin{table*}
\caption{A comparison of Equilibrium constants with \citet{as81} and \citet{h82} for the reaction \hp + HD $\rightarrow$ \hdp + \hh.}
\label{tab:as81_com}
\begin{tabular}{llllll}
\hline
\hline
T(K) & Adams and Smith  & Herbst & This work &\\ \hline 80 & 4.48
($\pm$1.3) & 5.9 & 6.82 & \\ 200 & 2.35 ($\pm$0.7) & 2.6 & 1.52 & \\
295 & 1.96 ($\pm$0.6) & 2.1$^a$ & 1.07$^a$ & \\ \hline
\multicolumn{3}{l}{\footnotesize{a. The theoretical value is actually at 300 K}}
\end{tabular}

\end{table*}

\begin{table*}
\caption{A comparison of Equilibrium constants at a temperature of 10 K for the reaction \hp + HD $\rightarrow$ \hdp + \hh.  Powers of ten given in parenthesis.
\label{tab:ghr02_com}
}
\begin{tabular}{ll}
\hline
\hline
This Work               & 2.6(+12) \\	
Gerlich \et\ \cite{ghr02} & 7.14 \\
Adams and Smith \cite{as81} & 1.5(+9)$^a$ \\
Sidhu \et \cite{smt92}   & 7.1(+12)$^b$ \\ \hline
\multicolumn{2}{l}{\footnotesize{a. This value is extrapolated from experimental data}}\\
\multicolumn{2}{l}{\footnotesize{b. This value uses the corrected enthalpy of 231.8 K}}
\end{tabular}
\end{table*}

\subsection{Transition Intensities}
Using wavefunctions produced in the calculation of the energy levels
outlined in section \ref{sec:energy_levels} and additional
wavefunctions calculated for \hdp\ in the same manner; the DIPOLE3
program of
\cite{jt334}; and  the dipole surface of \cite{rkw+94}; the transition
intensities were calculated. The dipole transition intensities for
\hdp, \dhp and \ddp are given in tables \ref{tab:h2dp_trans}, 
\ref{tab:d2hp_trans} and \ref{tab:d3p_trans} respectively;
transitions are given up to J=5 and a maximum frequency of 5000 \cm;
transitions whose relative intensity is less than 0.0001 are
neglected. The \ddp\ energy levels are labelled by the notation
($\nu_1$, $\nu_2$, J, G, U) \cite{wfm87}. The quantum numbers $\nu_1$,
$\nu_2$, G and U were assigned by referring to the work of
~\cite{acc94} and by inspection. The \hdp and \dhp levels are assigned
by hand using the standard quantum numbers J, K$_a$, and K$_c$.

\begin{table*}
{\begin{minipage}{210mm}
\caption{Einstein A coefficients for transitions form low-lying levels of \hdp. Powers of ten given in parenthesis.
\label{tab:h2dp_trans}}
\begin{tabular}{lllrrlllrrrrrr}
\hline
\hline
J$^{'}$ & K$^{'}_a$ & K$^{'}_c$ & & E$^{'}$ / \cm\  & J$^{''}$ & K$^{''}_a$ & K$^{''}_c$ & & E$^{''}$ / \cm\ & $\omega_{if}$(calc.) / \cm\ &  $\omega_{if}$(obs.) / \cm\ & A$_{if}$ / s$^{-1}$ \\
\hline
1 & 1 &  0 & B$_1$  &  72.457	 &  1 &  1 &  1  &    A$_1$  &   60.027     &   12.429	  & \multicolumn{1}{c}{-}		& 1.2186(-4)	\\
1 & 0 &  1 & A$_2$  &  45.698    &  0 &  0 &  0  &    A$_1$  &   0.000      & 	45.698	  & \multicolumn{1}{c}{-}     		& 4.0397(-3)	\\
2 & 1 &  2 & A$_2$  &  138.843   &  1 &  1 &  1  &    A$_1$  &   60.027     & 	78.816	  & \multicolumn{1}{c}{-}     		& 1.8762(-2)	\\
2 & 0 &  2 & A$_1$  &  131.638   &  1 &  0 &  1  &    A$_2$  &   45.698     & 	85.94	  & \multicolumn{1}{c}{-}     		& 3.0338(-2)	\\
2 & 1 &  1 & B$_1$  &  175.939   &  1 &  1 &  0  &    B$_1$  &   72.457     & 	103.483	  & \multicolumn{1}{c}{-}     		& 4.238(-2)	\\
2 & 2 &  0 & A$_1$  &  223.868   &  1 &  0 &  1  &    A$_2$  &   45.698     & 	178.17	  & \multicolumn{1}{c}{-}     		& 1.6643(-2)	\\
0 & 0 &  0 & A$_1$  &  2205.916  &  1 &  0 &  1  &    A$_2$  &   45.698     & 	2160.218  & 2160.176$^{a}$    	& 17.545	\\
1 & 1 &  0 & B$_1$  &  2278.465  &  1 &  1 &  1  &    A$_1$  &   60.027     & 	2218.438  & 2218.393$^{a}$    	& 10.372	\\
1 & 0 &  1 & A$_2$  &  2246.727  &  0 &  0 &  0  &    A$_1$  &   0.000      & 	2246.727  & 2246.697$^{a}$    	& 1.8962	\\
2 & 0 &  2 & A$_1$  &  2318.377  &  1 &  0 &  1  &    A$_2$  &   45.698     & 	2272.68	  & \multicolumn{1}{c}{-}     		& 0.35157	\\
0 & 0 &  0 & A$_2$  &  2335.338  &  1 &  1 &  1  &    A$_1$  &   60.027     & 	2275.31	  & 2275.403$^{a}$    	& 145.65	\\
1 & 0 &  1 & A$_1$  &  2383.878  &  1 &  1 &  0  &    B$_1$  &   72.457     & 	2311.421  & 2311.512$^{a}$    	& 83.419	\\
1 & 1 &  0 & B$_1$  &  2409.227  &  1 &  0 &  1  &    A$_2$  &   45.698     & 	2363.529  & \multicolumn{1}{c}{-}     		& 78.984	\\
2 & 2 &  0 & A$_1$  &  2427.119  &  1 &  0 &  1  &    A$_2$  &   45.698     & 	2381.421  & 2381.367$^{a}$    	& 3.058	\\
1 & 1 &  1 & A$_2$  &  2402.699  &  0 &  0 &  0  &    A$_1$  &   0.000      & 	2402.699  & 2402.795$^{a}$    	& 60.938	\\
2 & 0 &  2 & A$_2$  &  2477.681  &  1 &  1 &  1  &    A$_1$  &   60.027     & 	2417.654  & 2417.734$^{a}$    	& 29.82	\\
2 & 1 &  2 & A$_1$  &  2490.966  &  1 &  0 &  1  &    A$_2$  &   45.698     & 	2445.268  & 2445.348$^{a}$    	& 58.586	\\
2 & 2 &  1 & B$_1$  &  2568.382  &  1 &  1 &  0  &    B$_1$  &   72.457     & 	2495.925  & 2496.014$^{a}$    	& 60.067	\\
2 & 2 &  0 & A$_2$  &  2569.489  &  1 &  1 &  1  &    A$_1$  &   60.027     & 	2509.461  & 2509.541$^{a}$    	& 48.476	\\
0 & 0 &  0 & A$_1$  &  2992.524  &  1 &  0 &  1  &    A$_2$  &   45.698     & 	2946.826  & 2946.802$^{b}$    	& 53.167	\\
1 & 1 &  0 & B$_1$  &  3063.331  &  1 &  1 &  1  &    A$_1$  &   60.027     & 	3003.304  & 3003.276$^{b}$    	& 27.509	\\
1 & 0 &  1 & A$_2$  &  3038.198  &  0 &  0 &  0  &    A$_1$  &   0.000      & 	3038.198  & 3038.177$^{b}$    	& 20.353	\\
2 & 1 &  2 & A$_2$  &  3128.888  &  1 &  1 &  1  &    A$_1$  &   60.027     & 	3068.86	  & 3068.845$^{b}$    	& 20.088	\\
2 & 0 &  2 & A$_1$  &  3123.324  &  1 &  0 &  1  &    A$_2$  &   45.698     & 	3077.626  & 3077.611$^{b}$    	& 24.757	\\
2 & 1 &  1 & B$_1$  &  3167.147  &  1 &  1 &  0  &    B$_1$  &   72.457     & 	3094.69	  & 3094.671$^{b}$    	& 19.302	\\
2 & 2 &  0 & A$_1$  &  3209.847  &  1 &  0 &  1  &    A$_2$  &   45.698     & 	3164.149  & \multicolumn{1}{c}{-}     		& 1.5976	\\
0 & 0 &  0 & A$_1$  &  4287.61   &  1 &  0 &  1  &    A$_2$  &   45.698     &   4241.912  & \multicolumn{1}{c}{-}		& 16.953	\\
1 & 0 &  1 & A$_2$  &  4331.45   &  0 &  0 &  0  &    A$_1$  &   0.000      &   4331.45	  & \multicolumn{1}{c}{-}		& 9.6306	\\
2 & 1 &  2 & A$_2$  &  4412.461  &  1 &  1 &  1  &    A$_1$  &   60.027     & 	4352.434  & 4352.360$^{c}$    	& 14.871	\\
2 & 0 &  2 & A$_1$  &  4407.925  &  1 &  0 &  1  &    A$_2$  &   45.698     & 	4362.227  & \multicolumn{1}{c}{-}		& 15.45	\\
0 & 0 &  0 & A$_2$  &  4461.832  &  1 &  1 &  1  &    A$_1$  &   60.027     & 	4401.805  & \multicolumn{1}{c}{-}     		& 88.628	\\
2 & 1 &  1 & A$_2$  &  4512.486  &  1 &  0 &  1  &    A$_2$  &   45.698     & 	4466.788  & \multicolumn{1}{c}{-}     		& 0.41306	\\
1 & 1 &  0 & B$_1$  &  4536.348  &  1 &  0 &  1  &    A$_2$  &   45.698     & 	4490.65	  & \multicolumn{1}{c}{-}     		& 44.342	\\
2 & 0 &  2 & A$_2$  &  4555.88   &  1 &  1 &  1  &    A$_1$  &   60.027     &   4495.853  & 4495.881$^{c}$	& 27.013	\\
1 & 1 &  1 & A$_2$  &  4512.558  &  0 &  0 &  0  &    A$_1$  &   0.000      & 	4512.558  & 4512.567$^{c}$    	& 41.115	\\
2 & 1 &  2 & A$_1$  &  4563.405  &  1 &  0 &  1  &    A$_2$  &   45.698     & 	4517.707  & \multicolumn{1}{c}{-}     		& 38.458	\\
0 & 0 &  0 & A$_1$  &  4602.746  &  1 &  0 &  1  &    A$_2$  &   45.698     & 	4557.048  & \multicolumn{1}{c}{-}     		& 69.044	\\
2 & 2 &  1 & B$_1$  &  4677.548  &  1 &  1 &  0  &    B$_1$  &   72.457     & 	4605.091  & \multicolumn{1}{c}{-}     		& 38.831	\\
1 & 2 &  1 & B$_1$  &  4677.74   &  1 &  1 &  1  &    A$_1$  &   60.027     &   4617.713  & \multicolumn{1}{c}{-} 		& 33.563	\\
2 & 2 &  0 & A$_2$  &  4691.531  &  1 &  1 &  1  &    A$_1$  &   60.027     & 	4631.504  & \multicolumn{1}{c}{-}     		& 20.198	\\
1 & 0 &  1 & A$_2$  &  4657.859  &  0 &  0 &  0  &    A$_1$  &   0.000      & 	4657.859  & \multicolumn{1}{c}{-}     		& 9.0071	\\
2 & 0 &  2 & A$_1$  &  4761.399  &  1 &  0 &  1  &    A$_2$  &   45.698     & 	4715.701  & \multicolumn{1}{c}{-}     		& 7.7521	\\
2 & 2 &  0 & B$_1$  &  4845.211  &  1 &  0 &  1  &    A$_2$  &   45.698     & 	4799.514  & \multicolumn{1}{c}{-}     		& 1.3349	\\
0 & 0 &  0 & A$_1$  &  5039.84   &  1 &  0 &  1  &    A$_2$  &   45.698     &	4994.142  & \multicolumn{1}{c}{-} 		& 10.723	\\
\hline	
\hline
\multicolumn{12}{l}{\footnotesize{a. Frequencies of \cite{fmp86}}} \\
\multicolumn{12}{l}{\footnotesize{b. Frequencies of \cite{kpz88}}} \\
\multicolumn{12}{l}{\footnotesize{c. Frequencies of \cite{fd02}}} 

\end{tabular} 
\end{minipage}}
\end{table*}

\begin{table*}
{\begin{minipage}{210mm}
\caption{Einstein A coefficients for transitions form low-lying levels of \dhp. Powers of ten given in parenthesis.
\label{tab:d2hp_trans}}
\begin{tabular}{lllrrlllrrrrrr}
\hline
\hline
J$^{'}$ & K$^{'}_a$ & K$^{'}_c$ & & E$^{'}$ / \cm\ & J$^{''}$ & K$^{''}_a$ & K$^{''}_c$ & & E$^{''}$ / \cm\ & $\omega_{if}$(calc.) /  \cm\ &  $\omega_{if}$(obs.) / \cm\ & A$_{if}$ / s$^{-1}$ \\
\hline
1 & 1 & 0 & B$_{1}$ & 57.993   & 1 & 0 & 1 & A$_{1}$ & 34.918 &  23.075   & \multicolumn{1}{c}{-} 		&5.0911(-4) \\
1 & 1 & 1 & A$_{2}$ & 49.255   & 0 & 0 & 0 & A$_{1}$ & 0.000  &  49.255   & \multicolumn{1}{c}{-} 		&3.3031(-3)  \\
2 & 0 & 2 & A$_{1}$ & 101.716  & 1 & 1 & 1 & A$_{2}$ & 49.255 &  52.461   & \multicolumn{1}{c}{-} 		&2.0702(-3)  \\
2 & 1 & 2 & A$_{2}$ & 110.259  & 1 & 0 & 1 & A$_{1}$ & 34.918 &  75.341   & \multicolumn{1}{c}{-}      	&1.0597(-2)\\
2 & 2 & 1 & B$_{1}$ & 179.173  & 1 & 1 & 0 & B$_{1}$ & 57.993 &  121.18   & \multicolumn{1}{c}{-}     	&4.4504(-2)\\
2 & 2 & 0 & A$_{1}$ & 182.074  & 1 & 1 & 1 & A$_{2}$ & 49.255 &  132.819  & \multicolumn{1}{c}{-}		&4.4629(-2)\\
0 & 0 & 0 & A$_{1}$ & 1968.146 & 1 & 1 & 1 & A$_{2}$ & 49.255 &  1918.89  & 1918.908$^{a}$	&52.175	\\
1 & 0 & 1 & A$_{1}$ & 1998.523 & 1 & 1 & 0 & B$_{1}$ & 57.993 &  1940.53  & 1940.551$^{a}$	&21.989	\\
1 & 1 & 0 & B$_{1}$ & 2027.034 & 1 & 0 & 1 & A$_{1}$ & 34.918 &  1992.116 & 1992.130$^{a}$	&28.48	\\
2 & 0 & 2 & A$_{1}$ & 2055.077 & 1 & 1 & 1 & A$_{2}$ & 49.255 &  2005.821 & 2005.844$^{a}$ 	&6.0891	\\
1 & 1 & 1 & A$_{2}$ & 2014.09  & 0 & 0 & 0 & A$_{1}$ & 0.000  &  2014.09  & 2014.106$^{a}$	&14.364	\\
2 & 1 & 2 & A$_{2}$ & 2062.923 & 1 & 0 & 1 & A$_{1}$ & 34.918 &  2028.005 & 2028.024$^{a}$	&9.9809	\\
0 & 0 & 0 & A$_{2}$ & 2078.435 & 1 & 0 & 1 & A$_{1}$ & 34.918 &  2043.517 & 2043.515$^{a}$	&21.728	\\
1 & 1 & 1 & A$_{1}$ & 2128.7   & 1 & 1 & 0 & B$_{1}$ & 57.993 &  2070.707 & 2070.708$^{a}$	&16.782	\\
1 & 1 & 0 & A$_{2}$ & 2136.248 & 1 & 1 & 1 & A$_{2}$ & 49.255 &  2086.992 & 2086.990$^{a}$	&12.135	\\
2 & 2 & 1 & B$_{1}$ & 2145.612 & 1 & 1 & 0 & B$_{1}$ & 57.993 &  2087.619 & 2087.630$^{a}$	&11.233	\\
2 & 2 & 0 & A$_{1}$ & 2149.555 & 1 & 1 & 1 & A$_{2}$ & 49.255 &  2100.299 & 2100.307$^{a}$	&5.9919	\\
1 & 0 & 1 & A$_{2}$ & 2118.589 & 0 & 0 & 0 & A$_{1}$ & 0.000  &  2118.589 & 2118.588$^{a}$	&14.303	\\
2 & 1 & 2 & A$_{1}$ & 2202.779 & 1 & 1 & 1 & A$_{2}$ & 49.255 &  2153.524 & 2153.525$^{a}$	&21.901	\\
2 & 0 & 2 & A$_{2}$ & 2194.064 & 1 & 0 & 1 & A$_{1}$ & 34.918 &  2159.146 & 2159.145$^{a}$	&19.827	\\
2 & 1 & 1 & B$_{1}$ & 2225.161 & 1 & 1 & 0 & B$_{1}$ & 57.993 &  2167.168 & 2167.166$^{a}$	&17.293	\\
2 & 2 & 0 & A$_{2}$ & 2257.594 & 1 & 0 & 1 & A$_{1}$ & 34.918 &  2222.675 & \multicolumn{1}{c}{-} 		&0.38661\\
0 & 0 & 0 & A$_{1}$ & 2736.98  & 1 & 1 & 1 & A$_{2}$ & 49.255 &  2687.724 & \multicolumn{1}{c}{-} 		&86.183	\\
1 & 0 & 1 & A$_{1}$ & 2771.523 & 1 & 1 & 0 & B$_{1}$ & 57.993 &  2713.53  & \multicolumn{1}{c}{-} 		&44.856	\\
1 & 1 & 0 & B$_{1}$ & 2793.96  & 1 & 0 & 1 & A$_{1}$ & 34.918 &  2759.042 & 2759.036$^{b}$    &44.572	\\
1 & 1 & 1 & A$_{2}$ & 2785.338 & 0 & 0 & 0 & A$_{1}$ & 0.000  &  2785.338 & 2785.332$^{b}$    &31.182	\\
2 & 0 & 2 & A$_{1}$ & 2837.556 & 1 & 1 & 1 & A$_{2}$ & 49.255 &  2788.3   & 2788.300$^{b}$	&16.939	\\
2 & 1 & 2 & A$_{2}$ & 2845.72  & 1 & 0 & 1 & A$_{1}$ & 34.918 &  2810.802 & 2810.800$^{b}$ 	&29.412	\\
2 & 2 & 1 & B$_{1}$ & 2912.708 & 1 & 1 & 0 & B$_{1}$ & 57.993 &  2854.716 & 2854.707$^{b}$	&29.321	\\
2 & 2 & 0 & A$_{1}$ & 2915.616 & 1 & 1 & 1 & A$_{2}$ & 49.255 &  2866.36  & 2866.350$^{b}$	&22.136	\\
0 & 0 & 0 & A$_{1}$ & 3821.309 & 1 & 1 & 1 & A$_{2}$ & 49.255 &  3772.054 & \multicolumn{1}{c}{-}      	&6.5942	\\
1 & 0 & 1 & A$_{1}$ & 3851.977 & 1 & 1 & 0 & B$_{1}$ & 57.993 &  3793.984 & \multicolumn{1}{c}{-}     	&5.0848	\\
1 & 1 & 0 & B$_{1}$ & 3881.727 & 1 & 0 & 1 & A$_{1}$ & 34.918 &  3846.809 & 3846.786$^{c}$	&3.3258	\\
2 & 0 & 2 & A$_{1}$ & 3909.933 & 1 & 1 & 1 & A$_{2}$ & 49.255 &  3860.677 & 3860.660$^{c}$	&3.0475	\\
1 & 1 & 1 & A$_{2}$ & 3871.398 & 0 & 0 & 0 & A$_{1}$ & 0.000  &  3871.398 & 3871.377$^{c}$	&3.6663	\\
2 & 1 & 2 & A$_{2}$ & 3921.988 & 1 & 0 & 1 & A$_{1}$ & 34.918 &  3887.07  & 3887.052$^{c}$	&5.4261	\\
2 & 2 & 1 & A$_{1}$ & 4010.528 & 1 & 1 & 0 & B$_{1}$ & 57.993 &  3952.535 & \multicolumn{1}{c}{-}      	&3.6408	\\
2 & 2 & 0 & A$_{1}$ & 4013.18  & 1 & 1 & 1 & A$_{2}$ & 49.255 &  3963.924 & \multicolumn{1}{c}{-}     	&3.4514	\\
0 & 0 & 0 & A$_{1}$ & 4042.815 & 1 & 1 & 1 & A$_{2}$ & 49.255 &  3993.56  & 3993.518$^{c}$	&47.457	\\
1 & 0 & 1 & A$_{1}$ & 4058.521 & 1 & 1 & 0 & B$_{1}$ & 57.993 &  4000.528 & 4000.494$^{c}$ 	&42.9	\\
0 & 0 & 0 & A$_{2}$ & 4060.822 & 1 & 0 & 1 & A$_{1}$ & 34.918 &  4025.904 & 4025.873$^{c}$ 	&41.465	\\
2 & 0 & 2 & A$_{1}$ & 4097.094 & 1 & 1 & 1 & A$_{2}$ & 49.255 &  4047.839 & 4047.840$^{c}$ 	&21.275	\\
1 & 1 & 1 & A$_{2}$ & 4062.925 & 0 & 0 & 0 & A$_{1}$ & 0.000  &  4062.925 & 4062.889$^{c}$   	&29.138	\\
2 & 1 & 2 & A$_{1}$ & 4097.933 & 1 & 0 & 1 & A$_{1}$ & 34.918 &  4063.015 & 4062.983$^{c}$   	&26.675	\\
1 & 1 & 0 & A$_{2}$ & 4101.122 & 1 & 0 & 1 & A$_{1}$ & 34.918 &  4066.204 & 4066.158$^{c}$	&23.775	\\
1 & 0 & 1 & B$_{1}$ & 4119.147 & 1 & 1 & 1 & A$_{2}$ & 49.255 &  4069.891 & 4069.859$^{c}$ 	&20.300	\\
2 & 2 & 1 & B$_{1}$ & 4179.804 & 1 & 1 & 0 & B$_{1}$ & 57.993 &  4121.811 & \multicolumn{1}{c}{-} 		&24.474	\\
1 & 1 & 1 & A$_{2}$ & 4122.993 & 0 & 0 & 0 & A$_{1}$ & 0.000  &  4122.993 & \multicolumn{1}{c}{-} 		&1.3689(-3)\\
2 & 0 & 2 & A$_{1}$ & 4214.033 & 1 & 1 & 1 & A$_{2}$ & 49.255 &  4164.777 & \multicolumn{1}{c}{-}      	&5.7652	\\
2 & 2 & 0 & A$_{2}$ & 4208.006 & 1 & 0 & 1 & A$_{1}$ & 34.918 &  4173.088 & \multicolumn{1}{c}{-}     	&0.92222\\
2 & 1 & 2 & A$_{1}$ & 4229.853 & 1 & 1 & 1 & A$_{2}$ & 49.255 &  4180.597 & \multicolumn{1}{c}{-}		&2.4239	\\
2 & 2 & 1 & A$_{2}$ & 4252.4   & 1 & 0 & 1 & A$_{1}$ & 34.918 &  4217.482 & \multicolumn{1}{c}{-} 		&1.3913	\\
0 & 0 & 0 & A$_{1}$ & 4648.808 & 1 & 1 & 1 & A$_{2}$ & 49.255 &  4599.553 & \multicolumn{1}{c}{-} 		&10.723	\\
1 & 0 & 1 & A$_{1}$ & 4673.469 & 1 & 1 & 0 & B$_{1}$ & 57.993 &  4615.476 & \multicolumn{1}{c}{-} 		&9.1384	\\
0 & 0 & 0 & A$_{2}$ & 4674.96  & 1 & 0 & 1 & A$_{1}$ & 34.918 &  4640.042 & \multicolumn{1}{c}{-}      	&13.692	\\
2 & 0 & 2 & A$_{1}$ & 4720.562 & 1 & 1 & 1 & A$_{2}$ & 49.255 &  4671.307 & \multicolumn{1}{c}{-}     	&4.6985	\\
1 & 1 & 0 & B$_{1}$ & 4706.784 & 1 & 0 & 1 & A$_{1}$ & 34.918 &  4671.866 & \multicolumn{1}{c}{-}		&4.9443	\\
1 & 1 & 1 & A$_{2}$ & 4681.85  & 0 & 0 & 0 & A$_{1}$ & 0.000  &  4681.85  & \multicolumn{1}{c}{-} 		&7.1377	\\
1 & 1 & 0 & A$_{1}$ & 4732.173 & 1 & 1 & 1 & A$_{2}$ & 49.255 &  4682.918 & \multicolumn{1}{c}{-} 		&6.9989	\\
2 & 1 & 2 & A$_{2}$ & 4723.648 & 1 & 0 & 1 & A$_{1}$ & 34.918 &  4688.73  & \multicolumn{1}{c}{-} 		&6.6333	\\
1 & 0 & 1 & A$_{2}$ & 4727.065 & 0 & 0 & 0 & A$_{1}$ & 0.000  &  4727.065 & \multicolumn{1}{c}{-}      	&0.44733\\
2 & 2 & 1 & B$_{1}$ & 4796.924 & 1 & 1 & 0 & B$_{1}$ & 57.993 &  4738.931 & \multicolumn{1}{c}{-}     	&6.7834	\\
2 & 2 & 0 & A$_{1}$ & 4798.798 & 1 & 1 & 1 & A$_{2}$ & 49.255 &  4749.542 & \multicolumn{1}{c}{-}		&2.1769	\\
2 & 0 & 2 & A$_{2}$ & 4807.782 & 1 & 0 & 1 & A$_{1}$ & 34.918 &  4772.864 & \multicolumn{1}{c}{-}		&0.87466\\
2 & 0 & 0 & B$_{1}$ & 4852.274 & 1 & 0 & 1 & A$_{1}$ & 34.918 &  4817.356 & \multicolumn{1}{c}{-}		&0.50311\\
\hline
\hline
\multicolumn{12}{l}{\footnotesize{a. Frequencies of \cite{pm89}}} \\
\multicolumn{12}{l}{\footnotesize{b. Frequencies of \cite{kpz88}}} \\
\multicolumn{12}{l}{\footnotesize{c. Frequencies of \cite{fd02}}}
\end{tabular}  
\end{minipage}}
\end{table*}

\begin{table*}
{\begin{minipage}{210mm}
\caption{Einstein A coefficients for transitions form low-lying levels of \ddp
\label{tab:d3p_trans}}
\begin{tabular}{clrrlrclrrlrccr}
\hline
\hline
J$^{'}$ & $\nu_{1}^{'}\nu_{2}^{'}$ & G$^{'}$ & U$^{'}$ &  & E$^{'}$ / & J$^{'}$ & $\nu_{1}^{''}\nu_{2}^{''}$ & G$^{''}$ & U$^{''}$ & & E$^{''}$ / & $\omega_{if}$(calc.) / & $\omega_{if}$(obs.)$^a$ / & A$_{if}$ / \\
\multicolumn{5}{c}{} & \cm & \multicolumn{5}{c}{} & \cm &  \cm & \cm & $s^{1}$ \\ 
\hline
0 & 01 & 1  & 1  & E$^{'}$    & 1834.655 & 1 & 00 &  1 & 0 & E$^{''}$  & 32.324 &  1802.331  &   1802.349 & 264.18   \\
1 & 01 & 0  & -1 & A$_2^{''}$ & 1884.380 & 1 & 00 &  0 & 0 & A1$^{'}$  & 43.609 &  1840.770  &   1840.789 & 34.99    \\
1 & 01 & 1  & 1  & E$^{'}$    & 1878.561 & 1 & 00 &  1 & 0 & E$^{''}$  & 32.324 &  1846.237  &   1846.256 & 141.96   \\
1 & 01 & 0  & 1  & A$_1^{''}$ & 1888.048 & 0 & 00 &  0 & 0 & A$_1^{'}$ & 0.000  &  1888.048  &   1888.065 & 252.84   \\
2 & 01 & 1  & -1 & E$^{'}$    & 1955.981 & 1 & 00 &  1 & 0 & E$^{''}$  & 32.324 &  1923.657  &   1923.670 & 107.50   \\
2 & 01 & 0  & 1  & A$_2^{''}$ & 1979.203 & 1 & 00 &  0 & 0 & A$_2^{'}$ & 43.609 &  1935.593  &   1935.609 & 24.65    \\
2 & 01 & 1  & 1  & E$^{'}$    & 1968.077 & 1 & 00 &  1 & 0 & E$^{''}$  & 32.324 &  1935.753  & 	 \multicolumn{1}{c}{-}	  & 119.38   \\
1 & 02 & 3  & 2  & A$_2^{'}$  & 3646.285 & 1 & 00 &  0 & 0 & A$_1^{'}$ & 43.609 &  3602.676  &   3602.669 & 234.05   \\
0 & 02 & 2  & 2  & E$^{'}$    & 3650.700 & 1 & 00 &  1 & 0 & E$^{''}$  & 32.324 &  3618.376  &   3618.371 & 231.47   \\
2 & 02 & 4  & 2  & E$^{'}$    & 3662.342 & 1 & 00 &  1 & 0 & E$^{''}$  & 32.324 &  3630.018  &   3630.022 & 140.29   \\
1 & 02 & -3 & -2 & A$_1^{''}$ & 3647.190 & 0 & 00 &  0 & 0 & A$_1^{'}$ & 0.000  &  3647.190  & 	 \multicolumn{1}{c}{-}	  & 193.96   \\
1 & 02 & 2  & 2  & E$^{'}$    & 3694.893 & 1 & 00 &  1 & 0 & E$^{''}$  & 32.324 &  3662.569  &   3662.557 & 114.64   \\
2 & 02 & -3 & 2  & A$_2^{''}$ & 3736.400 & 1 & 00 &  0 & 0 & A$_2^{'}$ & 43.609 &  3692.791  &	 3692.785 & 17.29    \\
2 & 02 & 2  & 2  & E$^{'}$    & 3783.234 & 1 & 00 &  1 & 0 & E$^{''}$  & 32.324 &  3750.909  &   3750.879 & 23.66    \\
\hline
\hline
\multicolumn{15}{l}{\footnotesize{a. Frequencies of Amano \et\ \cite{acc94}}}
\end{tabular} 
\end{minipage}}
\end{table*}

\section{Conclusions}
Partition functions for \hp\, \dhp\ and \ddp\ have been
calculated. These have been used to calculate the equilibrium
constants for the important gas phase reactions involving \hp\ and the
deuterated isotopomers. In general {\it ab initio} calculations are
easier to perform at very low temperatures because they require few
levels whereas these very low temperature experiments are extremely
challenging. In addition transition data have been calculated which
should aid in the observation of deuterated isotopomers of \hp. It is
hoped our data facilitate the understanding of deuterium chemistry in
the interstellar medium.

\section*{Acknowledgments}
We thank Brian Sutcliffe for helpful discussions. This work was
carried out on the Ra Supercomputer, at the HiPerSPACE Computing
Centre, UCL

\bibliographystyle{/amp/tex/styles/MNRAS/mn2e}

\begin{thebibliography}{}

\bibitem[\protect\citeauthoryear{{Adams} \& {Smith}}{{Adams} \&
  {Smith}}{1981}]{as81}
{Adams} N.~G.,  {Smith} D.,  1981, Astrophysical Journal, 248, 373

\bibitem[\protect\citeauthoryear{Amano, Chan, Civi\v{s}, McKellar, Majewski,
  Sadovsi\'{i} \& Watson}{Amano et~al.}{1994}]{acc94}
Amano T.,  Chan M.,  Civi\v{s} S.,  McKellar A.,  Majewski W.,  Sadovsi\'{i}
  D.,    Watson J.,  1994, Can J. Phys., 72, 1007

\bibitem[\protect\citeauthoryear{{Bacmann}, {Lefloch}, {Ceccarelli},
  {Steinacker}, {Castets} \& {Loinard}}{{Bacmann} et~al.}{2003}]{blc03}
{Bacmann} A.,  {Lefloch} B.,  {Ceccarelli} C.,  {Steinacker} J.,  {Castets} A.,
     {Loinard} L.,  2003, Astrophysical Journal, 585, L55

\bibitem[\protect\citeauthoryear{Carney}{Carney}{1980}]{car80}
Carney G.~D.,  1980, Chem. Phys., 54

\bibitem[\protect\citeauthoryear{{Caselli}, {van der Tak}, {Ceccarelli} \&
  {Bacmann}}{{Caselli} et~al.}{2003}]{ctc03}
{Caselli} P.,  {van der Tak} F.~F.~S.,  {Ceccarelli} C.,    {Bacmann} A.,
  2003, Astro. Astrophys., 403, L37

\bibitem[\protect\citeauthoryear{Ceccarelli}{Ceccarelli}{2002}]{c02}
Ceccarelli C.,  2002, Planet Space Sci., 50, 1267

\bibitem[\protect\citeauthoryear{Cencek, Rychlewski, Jaquet \&
  Kutzelnigg}{Cencek et~al.}{1998}]{crj98}
Cencek W.,  Rychlewski J.,  Jaquet R.,    Kutzelnigg W.,  1998, J. Chem. Phys.,
  108, 2831

\bibitem[\protect\citeauthoryear{Dalgarno \& Lepp}{Dalgarno \&
  Lepp}{1984}]{dl84}
Dalgarno A.,  Lepp S.,  1984, Astrophysical Journal, 287, L47

\bibitem[\protect\citeauthoryear{F\'{a}rn\'{i}k, Davis, Kostin, Polyansky,
  Tennyson \& Nesbitt}{F\'{a}rn\'{i}k et~al.}{2002}]{fd02}
F\'{a}rn\'{i}k M.,  Davis S.,  Kostin M.~A.,  Polyansky O.~L.,  Tennyson J.,
  Nesbitt D.~J.,  2002, J. Chem. Phys., 116, 6146

\bibitem[\protect\citeauthoryear{{Foster}, {McKellar}, {Peterkin}, {Watson} \&
  {Pan}}{{Foster} et~al.}{1986}]{fmp86}
{Foster} S.~C.,  {McKellar} A.~R.~W.,  {Peterkin} I.~R.,  {Watson} J.~K.~G.,
  {Pan} F.~S.,  1986, J. Chem. Phys., 84, 91

\bibitem[\protect\citeauthoryear{{Gerlich}, {Herbst} \& {Roueff}}{{Gerlich}
  et~al.}{2002}]{ghr02}
{Gerlich} D.,  {Herbst} E.,    {Roueff} E.,  2002, Planetary and Space Science,
  50, 1275

\bibitem[\protect\citeauthoryear{Giles, Adams,  \& Smith}{Giles
  et~al.}{1992}]{gas92}
Giles K.,  Adams N.~G.,     Smith D.,  1992, Journal of Physical Chemistry, 96,
  7645

\bibitem[\protect\citeauthoryear{Hayman}{Hayman}{1967}]{h67}
Hayman H.,  1967, Statistical thermodynamics : an introduction to its
  foundations.
Elsevier

\bibitem[\protect\citeauthoryear{{Herbst}}{{Herbst}}{1982}]{h82}
{Herbst} E.,  1982, Astro. Astrophys., 111, 76

\bibitem[\protect\citeauthoryear{Huber \& Herzberg}{Huber \&
  Herzberg}{1979}]{hh79}
Huber K.,  Herzberg G.,  1979, Constants of Diatomic Molecules.
Van Nostrand Reinhold Company

\bibitem[\protect\citeauthoryear{Kelly}{Kelly}{1987}]{k87}
Kelly R.,  1987, J. Phys. Chem. Ref. Data, 16

\bibitem[\protect\citeauthoryear{{Kozin}, {Polyansky} \& {Zobov}}{{Kozin}
  et~al.}{1988}]{kpz88}
{Kozin} I.~N.,  {Polyansky} O.~L.,    {Zobov} N.~F.,  1988, Journal of
  Molecular Spectroscopy, 128, 126

\bibitem[\protect\citeauthoryear{{Lemoine}, {Audouze}, {Ben Jaffel}, {Feldman},
  {Ferlet}, {Hebrard}, {Jenkins}, {Mallouris}, {Moos}, {Sembach}, {Sonneborn},
  {Vidal-Madjar} \& {York}}{{Lemoine} et~al.}{1999}]{laj99}
{Lemoine} M.,  {Audouze} J.,  {Ben Jaffel} L.,  {Feldman} P.,  {Ferlet} R.,
  {Hebrard} G.,  {Jenkins} E.~B.,  {Mallouris} C.,  {Moos} W.,  {Sembach} K.,
  {Sonneborn} G.,  {Vidal-Madjar} A.,    {York} D.~G.,  1999, New Astronomy, 4,
  231

\bibitem[\protect\citeauthoryear{Loinard, Castets, Ceccarelli, Caux \&
  Tielens}{Loinard et~al.}{2001}]{lcc01}
Loinard L.,  Castets A.,  Ceccarelli C.,  Caux E.,    Tielens A. G. G.~M.,
  2001, Astrophys. J.

\bibitem[\protect\citeauthoryear{Neale \& Tennyson}{Neale \&
  Tennyson}{1995}]{jt169}
Neale L.,  Tennyson J.,  1995, Astrophys. J., 454, L169

\bibitem[\protect\citeauthoryear{{Oka}}{{Oka}}{1980}]{o80}
{Oka} T.,  1980, Physical Review Letters, 45, 531

\bibitem[\protect\citeauthoryear{{Oka}}{{Oka}}{1981}]{o81}
{Oka} T.,  1981, Royal Society of London Philosophical Transactions Series A,
  303, 543

\bibitem[\protect\citeauthoryear{Polyansky \& McKellar}{Polyansky \&
  McKellar}{1989}]{pm89}
Polyansky O.,  McKellar A.,  1989, J. Chem. Phys., 92, 4039

\bibitem[\protect\citeauthoryear{Polyansky \& Tennyson}{Polyansky \&
  Tennyson}{1999}]{jt236}
Polyansky O.~L.,  Tennyson J.,  1999, J. Chem. Phys., 110, 5056

\bibitem[\protect\citeauthoryear{Ramanlal, Polyansky \& Tennyson}{Ramanlal
  et~al.}{2003}]{jt318}
Ramanlal J.,  Polyansky O.~L.,    Tennyson J.,  2003, Astron. Astrophys., 406,
  383

\bibitem[\protect\citeauthoryear{Roberts, Herbst \& Millar}{Roberts
  et~al.}{2003}]{rhm03}
Roberts H.,  Herbst E.,    Millar T.~J.,  2003, Astrophysical Journal, 591, L41

\bibitem[\protect\citeauthoryear{{R\"{o}hse}, Kutzelnigg, Jaquet \&
  Klopper}{{R\"{o}hse} et~al.}{1994}]{rkw+94}
{R\"{o}hse} R.,  Kutzelnigg W.,  Jaquet R.,    Klopper W.,  1994, J. Chem.
  Phys., 101, 2231

\bibitem[\protect\citeauthoryear{Sauval \& Tatum}{Sauval \& Tatum}{1984}]{st84}
Sauval A.~J.,  Tatum J.~B.,  1984, Astrophysical Journal Supplement Series, 56,
  193

\bibitem[\protect\citeauthoryear{Sidhu, Miller \& Tennyson}{Sidhu
  et~al.}{1992}]{jt113}
Sidhu K.~S.,  Miller S.,    Tennyson J.,  1992, Astron. Astrophys., 255, 453

\bibitem[\protect\citeauthoryear{Smith, Adams \& Alge}{Smith
  et~al.}{1982}]{saa82}
Smith D.,  Adams N.~G.,    Alge E.,  1982, Astrophysical Journal, 263, 123

\bibitem[\protect\citeauthoryear{Stark, van~der Tak \& van Dishoeck}{Stark
  et~al.}{1999}]{std99}
Stark R.,  van~der Tak F. F.~S.,    van Dishoeck E.~F.,  1999, Astrophys. J.

\bibitem[\protect\citeauthoryear{Tennyson, Kostin, Barletta, Harris, Ramanlal,
  Polyansky \& Zobov}{Tennyson et~al.}{2004}]{jt334}
Tennyson J.,  Kostin M.~A.,  Barletta P.,  Harris G.~J.,  Ramanlal J.,
  Polyansky O.~L.,    Zobov N.~F.,  2004, Computer Phys. Comm.

\bibitem[\protect\citeauthoryear{{Vastel}, {Phillips} \& {Yoshida}}{{Vastel}
  et~al.}{2004}]{vpy04}
{Vastel} C.,  {Phillips} T.~G.,    {Yoshida} H.,  2004, Astrophysical Journal,
  606, L127

\bibitem[\protect\citeauthoryear{Watson, Foster \& Mckellar}{Watson
  et~al.}{1987}]{wfm87}
Watson J.,  Foster S.,    Mckellar A.,  1987, Can. J. Phys., 65, 38

\end{thebibliography}

\end{document}